\documentclass[a4paper,11pt]{article}
\usepackage{pos}
\usepackage{cleveref}
\usepackage{lineno}

\title{Radiation of ionisation electrons: the key role of their 2-pt function of velocities}

\author*[]{Olivier Deligny}

\affiliation[]{Laboratoire de Physique des 2 Infinis Ir\`ene Joliot-Curie (IJCLab)\\
CNRS/IN2P3, Universit\'{e} Paris-Saclay, Orsay, France}

\emailAdd{deligny@ijclab.in2p3.fr}

\abstract{
Several attempts to detect extensive air showers (EAS) induced by ultra-high energy cosmic rays have been conducted in the last decade based on the molecular Bremsstrahlung radiation (MBR) at GHz frequencies from quasi-elastic collisions of ionisation electrons left in the atmosphere after the passage of the cascade of particles. These attempts have led to the detection of a handful of signals only, all of them forward-directed along the shower axis and hence suggestive of originating from geomagnetic and Askaryan emissions extending into GHz frequencies close to the Cherenkov angle. In this contribution to ARENA2022, the lack of detection of events is explained by the coherent suppression of the MBR in frequency ranges below the collision rate due to the destructive interference impacting the emission amplitude of photons between the successive collisions of the electrons. The spectral intensity at the ground level is shown to be several orders of magnitude below the sensitivity of experimental setups. Consequently the MBR cannot be seen as the basis of a new detection technique of EAS for the next decades. The formalism developed to get at this conclusion allowed the key role of the two-point correlation function of the ionisation electron velocities to be highlighted. This can serve to study the intensity of the re-radiation of these ionization electrons subject to the passage of an incoming coherent wave from a radar transmitter. Some hints on this will be presented.
}

\FullConference{%
  9th International Workshop on Acoustic and Radio EeV Neutrino Detection Activities - ARENA2022\\
  7-10 June 2022\\
  Santiago de Compostela, Spain}

\newcommand{\dif}{\mathrm{d}}

\begin{document}
\maketitle


\section{Introduction}

The energy of an Extensive Air Shower (EAS) is deposited mainly by ionisation in the atmosphere. The resulting numerous ionisation electrons can excite nitrogen molecules, which subsequently radiate isotropically in the ultra-violet band while de-exciting. The detection of this fluorescence light with telescope stations made up of  photomultiplier tubes each monitoring a small portion of the sky was a breakthrough in the technique of collecting EAS~\cite{Bergeson:1977nw}. Thanks to the isotropic emission, EAS are observed side-on and their longitudinal profiles can be accurately reconstructed. Large detection areas can be covered by means of a few fluorescence stations only, spaced every 20 km or so. The downside of the technique is, however, its low duty cycle, about 10\%, due to the need for operating during moonless nights only. 
 
A giant step forward towards a future ground-based observatory providing a jump in the collection of ultra-high energy cosmic rays requires the development of a detection technique with stations widely spaced, ideally every 10--20~km as in the case of fluorescence telescope stations, and with high duty cycle. In this context, the various possible emissions from the plasma made of ionisation electrons and neutral molecules may provide viable mechanisms for the detection of EAS. 

One emission mechanism stems from the Bremsstrahlung radiation induced by the quasi-elastic scattering of ionisation electrons with molecular nitrogen and oxygen. This molecular Bremsstrahlung radiation (MBR) is isotropic and unpolarised. Triggered by microwave emission measurements in laboratory~\cite{Gorham:2007af}, MBR has been used in the last decade in the GHz band, in which radiation propagates in the atmosphere in a quasi-unattenuated way (less than 0.05~dB~km$^{-1}$), for performing shower calorimetry in the same spirit as the fluorescence technique does, by mapping the ionisation content along the showers through the intensity of the microwave signals detected at ground level. However, only a few handful signals forward-directed along the shower axes were recorded, with in particular no side-on observation of EAS~\cite{PierreAuger:2013qtf,Smida:2014sia}. One aim of this contribution is, following~\cite{Deligny:2021brf}, to explain the reasons of the faintness of the emission.

Another emission mechanism stems from the coherent re-radiation of the ionisation electrons subject to an incoming low-frequency wave. 
Radar detection of EAS is an old idea initially suggested in 1941 when reflections were proposed as a possible explanation for anomalous atmospheric radar echoes. This old idea has been updated in the last decade~\cite{Gorham:2000da}. At the Telescope Array, some tests have been pursued  using this technique~\cite{Abbasi:2016ugu}. The principle relies on considering the front of the EAS as a hard target able to reflect the incoming wave. The ``chirp'' signal, resulting from the motion of the front, is then expected at the reception station. No EAS could be, however, detected, yielding to constraints on the ``radar cross section'' excluding the viability of the technique as presented in~\cite{Gorham:2000da}. 

In the estimates presented in~\cite{Gorham:2000da}, the front of the EAS is considered as a conductor reflecting the incoming wave. However, the conditions of reflections, governed by requiring the continuity of the tangential component of the electric field between the air and the front of the shower, are not good enough to allow for the formation of a chirp signal at the reception station, because the mean current left just behind the shower front is annihilated almost instantaneously by the high (elastic) collision rate of the ionisation electrons~\cite{Stasielak:2014ewa}. This effect was neglected in~\cite{Gorham:2000da}. The incoming wave is thus penetrating into the plasma of ionisation electrons and molecules in air. Consequently, the relevant question is to know whether or not the incoming wave can be re-radiated by the ionisation electrons  so that a signal could be detected. Some exploratory hints on how to tackle this problem are given; they might be an helpful basis to interpret the signal in ice reported in~\cite{Prohira:2019glh}.

\section{Molecular Bremsstrahlung Radiation of Low-Energy Electrons in Extensive Air Showers}
\label{sec:mbr}

Let $n_{\mathrm{EAS}}$ be the number of high-energy charged particles per surface unit in an EAS and $\rho(\mathbf{x})$ the density of molecular nitrogen or oxygen in the atmosphere at the position $\mathbf{x}$. These high-energy electrons/positrons from the cascade are refereed to as ``primary electrons'' hereafter, in contrast to the ionisation electrons, the production of which per unit volume, per velocity band and per time unit follows from 
\begin{equation}
\label{eqn:nei}
n(\mathbf{x},\mathbf{v}_0,t_0)=\frac{\rho(\mathbf{x})f(\mathbf{v}_0,t_0)}{I_0+\left\langle T\right\rangle}~\left\langle\frac{\dif E}{\dif X}\right\rangle~n_{\mathrm{EAS}}(\mathbf{x}).
\end{equation}
Here, $I_0$ is the ionisation potential to create an electron-ion pair in air, the bracketed expression $\left\langle\dif E/\dif X\right\rangle$ stands for the mean energy loss of the EAS charged particles per grammage unit, and $f(\mathbf{v}_0,t_0)$ is the distribution in velocity and time of the resulting ionisation electrons, which is related to that expressed in terms of kinetic energy, $f_T\left(T\right)$, through the Jacobian transformation
\begin{equation}
\label{eqn:f}
f(\mathbf{v}_0,t_0)=\frac{mv_0}{4\pi\left(1-\left(v_0/c\right)^2\right)^{3/2}}f_T\left(T\left(v_0\right),t_0\right),
\end{equation}
with $m$ the mass of the electron. For primary charged particles in the cascade with $\geq~$MeV energies, an expression for the distribution $f_T\left(T,t_0=0\right)$ that accounts for relativistic effects as well as indistinguishability between primary and secondary electrons was derived in~\cite{Arqueros:2009zz} and is used here. In the energy range of interest, this expression leads to $\left\langle T\right\rangle\simeq 40~$eV, in agreement with the well-known stopping power. The remaining time dependence in $t_0$, reflecting the subsequent cascade of ionisation electrons produced by secondary electrons themselves as long as their kinetic energy is above $I_0$, is derived by Monte-Carlo below. 

As long as they remain free, ionisation electrons with density $n'\equiv n(\mathbf{x}',\mathbf{v}'_0,t'_0)$ can thus produce photons through the process of quasi-elastic collisions with neutral molecules in the atmosphere. For non-relativistic particles, the spectral radiated energy can be shown to read as~\cite{Deligny:2021brf}
\begin{equation}
\label{eqn:Eomega_final}
    \mathcal{E}(\omega)=\frac{e^2\omega^2}{6\pi^2\epsilon_0c^3}\iint\dif \mathbf{x'}\dif \mathbf{x''}\iint\dif \mathbf{v}'_0\dif \mathbf{v}''_0\iint\dif t'_0\dif t''_0\int_{t'_0}^{\infty}\int_{t''_0}^{\infty}\dif t'\dif t''~\left\langle \left(n'\mathbf{v}(t')\right)\cdot\left(n''\mathbf{v}(t'')\right)\right\rangle e^{-i\omega (t'-t'')},
\end{equation}
where $e$ is the electric charge, $\epsilon_0$ is the vacuum permittivity, $c$ is the speed of light, and $\mathbf{v}(t')$ is the electron velocity at retarded time $t'$. The use of the $\langle\cdot\rangle$ symbol stands for the average over realisations of the stochastic process that governs the dynamics of $\mathbf{v}(t')$. An electron appearing free at $t'=0$ and disappearing (by attachment) at $t'=\tau$ experiences accelerations during each collision. Here, the 2-pt correlation function of the electron velocities must account for the density of particles. For an incoherent process between independent particles such as the MBR, this 2-pt correlation function is diagonal in every variable governing the densities:
\begin{equation}
\label{eqn:2pt}
    \left\langle \left(n'\mathbf{v}(t')\right)\cdot\left(n''\mathbf{v}(t'')\right)\right\rangle=n'\delta(\mathbf{x}',\mathbf{x}'')\delta(\mathbf{v}'_0,\mathbf{v}''_0)\delta(t'_0,t''_0)~\left\langle \mathbf{v}(t')\cdot\mathbf{v}(t'')\right\rangle. 
\end{equation}
In this way, the radiation scales with the number of particles once integrating \cref{eqn:Eomega_final} over positions, initial velocities and initial creation time. 

The radiation is thus determined by the 2-pt correlation function of the velocities of a single electron, obtained by Monte-Carlo by simulating a large number of test particles with initial velocities drawn at random from $f$ and undergoing collisions. The main features of the different rates depend on the energy. The total momentum transfer collision rate goes from $\simeq 100~$GHz up to $\simeq 10~$THz in the explored kinetic energy range, with different inelastic contributions that can be found elsewhere~\cite{Arqueros:2009zz,Deligny:2021brf}. The tabulated 2-pt correlation function of the velocities of a single electron obtained from the Monte-Carlo simulation allows the determination of the spectrum of radiation sought for. The radiation is observed to be suppressed below THz frequencies, with a quadratic dependence in frequency in the GHz range. 
 
\begin{figure}[!t]
\centering
\includegraphics[width=10cm]{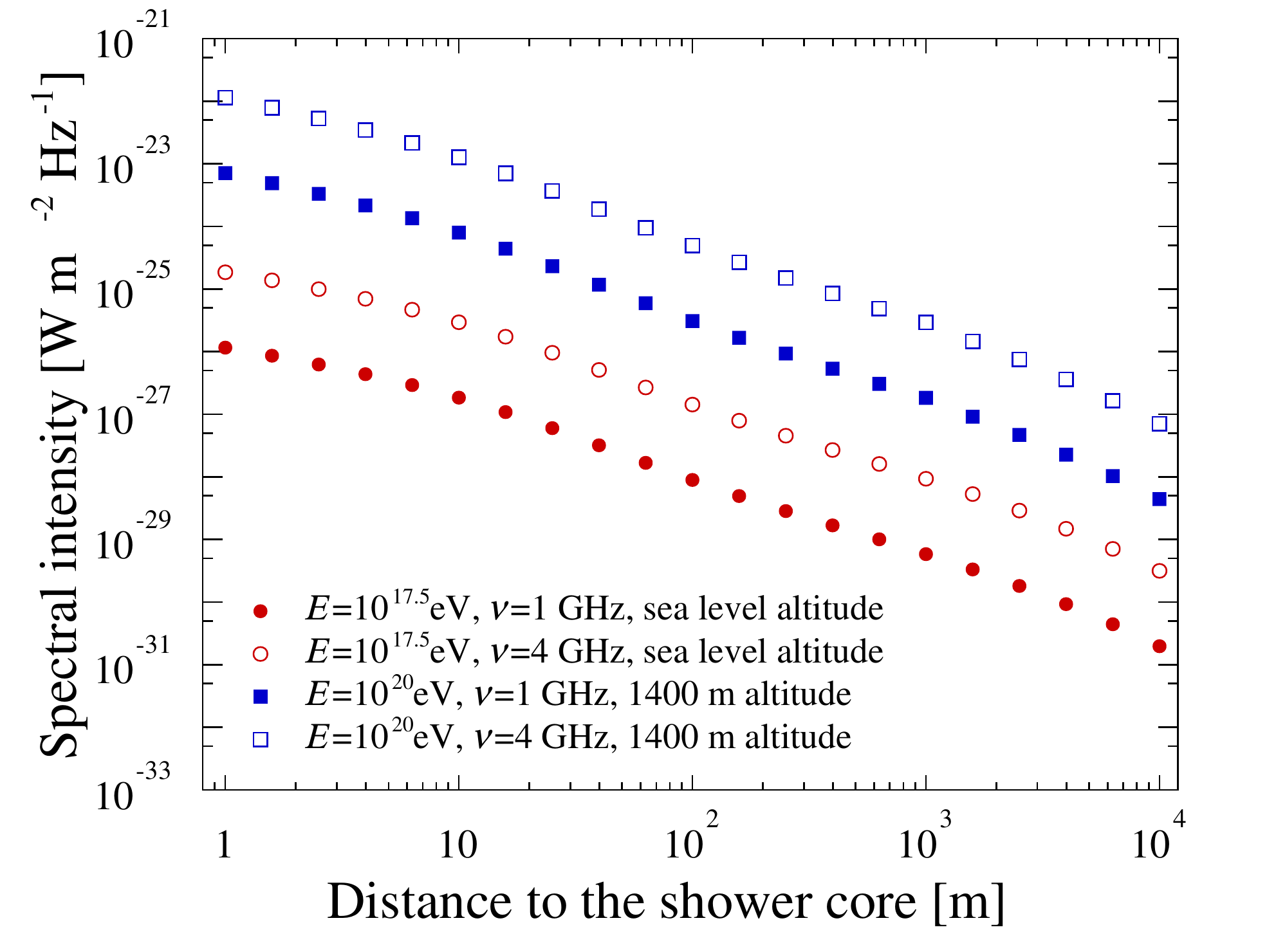}
\caption{\small{Spectral intensity from MBR as a function of the distance to the shower core. From~\cite{Deligny:2021brf}.}}
\label{fig:spectral_intensity}
\end{figure}

The spectral intensity, $\Phi(\nu,\mathbf{x}_{\mathrm{obs}})$, describes the power of the radiation received per unit frequency and passing through any unit area at an observation point $\mathbf{x}_{\mathrm{obs}}$. It results from the summation of the radiation emitted by all the ionisation electrons produced along the shower track. To provide relevant orders of magnitude for the spectral intensities that can be expected from MBR in the GHz band at the ground level, a crude model of the electromagnetic cascade of an EAS consisting of a longitudinal development described by a Gaisser-Hillas function and of a lateral extension described by an NKG function, limited to a vertical incidence, is used to infer an expression of $n_{\mathrm{EAS}}(\mathbf{x})$ to be plugged into \cref{eqn:nei}. The spectral intensity then results from 
\begin{equation}
\label{eqn:Phi}
\Phi(\nu,\mathbf{x}_{\mathrm{obs}})=\iiint \frac{r\dif r\dif\varphi\dif a}{4\pi R^2(r,\varphi,a)}~\mathcal{P}(\nu,a),
\end{equation}
where $\mathcal{P}(\nu,a)$ is the frequency spectrum of emitted power, obtained by dividing the radiated energy by the mean duration of the emission identified as the ``mean lifetime of the plasma'', which amounts, from the simulations described in the previous section, to $\simeq 30~$ns at the ground level.
 
The spectral intensity expected at different distances from the shower core is shown in \cref{fig:spectral_intensity}, for two primary energies and two frequencies of interest. The rapid decrease in amplitude for increasing distances is striking. Published limits on the MBR emission are currently at the level of $10^{-14.5}$~W~m$^{-2}$~Hz$^{-1}$~\cite{Alvarez-Muniz:2012ets}. The values derived in this study are by orders of magnitude below these current limits. A relevant estimate of the minimal spectral intensity $\Phi_{\mathrm{min}}$ detectable by an antenna operating in a bandwidth $\Delta\nu$ with a noise temperature $T_{\mathrm{sys}}$ and an effective area $A_{\mathrm{eff}}$ is known to obey
\begin{equation}
\label{eqn:Phimin}
\Phi_{\mathrm{min}}=\frac{kT_{\mathrm{sys}}}{A_{\mathrm{eff}}\sqrt{\tau\Delta\nu}},
\end{equation}
where $k$ is the Boltzmann constant and $\tau$ the receiver sampling time. For values $\Delta\nu=0.8$~GHz, $\tau=10~$ns, $T_{\mathrm{sys}}=50~$K and $A_{\mathrm{eff}}$ approaching $10^3~$cm$^2$, values typical of the setups used at the Pierre Auger Observatory, for instance, one gets $\Phi_{\mathrm{min}}$ on the order of a few $10^{-21}$~W~m$ ^{-2}$~Hz$^{-1}$. The results presented here show that the expected signals are out of reach of the experimental setups, even for a $10^{20}$~eV shower sampled at 1400~m altitude level, that of the Pierre Auger Observatory.  The spectral intensities turn out to be 7-to-8 orders of magnitude below the reference values anticipated from a scaling law converting the laboratory measurement to EAS expectations put forward in~\cite{Gorham:2007af}.

\section{Echo Radar from the Re-radiation of the Incoming Wave by Ionisation Electrons}
\label{sec:radar}

\begin{figure}[!t]  
\centering  \includegraphics[width=10cm]{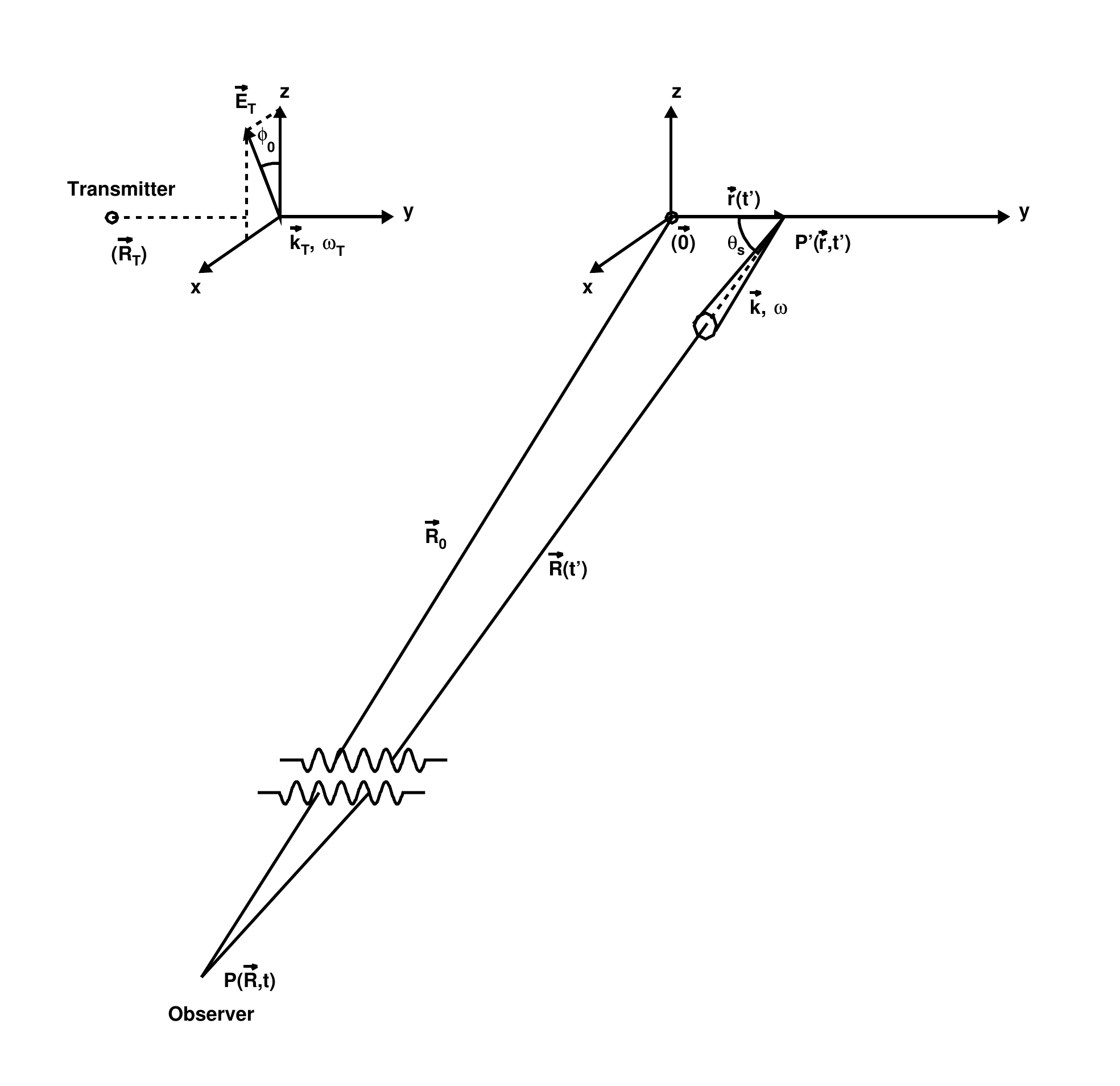}
\caption{\small{Geometry for the Thomson scattering. The accelerated electron is located in $P'$. $t'$ denotes the retarded time for the observer located in $P$.}}  
\label{fig:geom-thomson}
\end{figure}

We now turn into the question of the echo radar expected from the re-radiation of the incoming wave by ionisation electrons, providing some hints in the simple case of a plasma with uniform density. Referring to the geometry depicted in~\cref{fig:geom-thomson}, we start from the radiated energy $\mathcal{E}(\Omega)$ per unit solid angle flowing into an elementary cone $\dif\Omega$ and received at the observation point $P$ assumed to be far away from the accelerated charges located in $P'$:
\begin{equation}
\label{eqn:Eomega0}
    \mathcal{E}(\Omega)=\int\dif t~R^2\left\langle\mathbf{E}(R,t')\times\mathbf{H}^\star(R,t')\right\rangle\cdot\mathbf{q},
\end{equation}
where the radiated $\mathbf{E}$ and $\mathbf{H}$ fields depend on the retarded time $t'$, and $\mathbf{q}$ a unit vector in the observer direction that changes negligibly during a small acceleration interval. For $N$ non-relativistic particles, the approximation $\dif t'\simeq \dif t$ holds and the energy reads as
\begin{equation}
\label{Eomega1}
    \mathcal{E}(\Omega)=\frac{e^2}{16\pi^2\epsilon_0c^3}\left\langle\int_0^{T_p}\dif t'~\left|\sum_{p=1}^N\mathbf{q}\times\left(\mathbf{q}\times\dot{\mathbf{v}}_p(t')\right)\right|^2\right\rangle.
\end{equation}
From Parseval's identity, the spectral energy density per solid angle unit is then
\begin{equation}
\label{eqn:Enu}
    \mathcal{E}_\nu(\Omega)=\frac{e^2}{16\pi^3\epsilon_0c^3}\bigg\langle\bigg|\sum_{p=1}^N\int_0^{T_p}\hspace{-3mm}\dif t'\bigg(\mathbf{q}\times\bigg(\mathbf{q}\times\dot{\mathbf{v}}_p(t')\bigg)\bigg)\exp{\bigg(-i2\pi\nu t'-i\mathbf{k}\cdot\int_0^{\mathbf{R}_{0p}}\hspace{-3mm}n(\mathbf{r}')\dif\mathbf{r}'+i\mathbf{k}\cdot\int_0^{\mathbf{r}_p(t')}\hspace{-5mm}n(\mathbf{r}')\dif\mathbf{r}'\bigg)}\bigg|^2\bigg\rangle,
\end{equation}
where $\mathbf{R}(t')$ is approximated by $\mathbf{R}_0-\mathbf{r}(t')$, $\mathbf{k}=2\pi\nu\mathbf{q}/c$ is the wave vector, and $n(\mathbf{r}')$ is the altitude-dependent refractive index of the air.

A single electron appearing at $t'=0$ and disappearing at $t'=T_p$ (by attachment) experiences accelerations during each elastic collisions and under the influence of the incident electromagnetic wave of the radar whose electric vector is taken as
\begin{equation}
\label{eqn:ET}
    \mathbf{E}_{\mathrm{T}}\simeq\frac{U_{\mathrm{T}}}{R_{\mathrm{T}}}\exp\bigg(i2\pi\nu_{\mathrm{T}}t'+i\mathbf{k}_{\mathrm{T}}\cdot\int^{\mathbf{R}_{\mathrm{T},p}}_{0} \hspace{-3mm}n(\mathbf{r}')\mathrm{d}\mathbf{r}'-i\mathbf{k}_{\mathrm{T}}\cdot\int^{\mathbf{r}_p(t')}_{0} \hspace{-5mm}n(\mathbf{r}')\mathrm{d}\mathbf{r}'\bigg)\mathbf{n}_{\mathrm{T},\perp}.
\end{equation}
Here, $\omega_{\mathrm{T}}$ is the pulsation of the incident wave, $\mathbf{R}_{\mathrm{T}}$ the position of the transmitter, $\mathbf{r}_p$ the position of the electron, $\mathbf{k}_{\mathrm{T}}$ the wave vector oriented from the transmitter to the electron, and $\mathbf{n}_{\mathrm{T},\perp}$ a unit vector perpendicular to $\mathbf{k}_{\mathrm{T}}$. Note that the random phase of the emitter is ignored hereafter, since it plays no logical role. In practical cases considered here, the voltage amplitude $U_{\mathrm{T}}$ is sufficiently small so that the accelerated charges do not attain relativistic velocities. The acceleration due to the radar wave is simply $e\mathbf{E}_{\mathrm{T}}/m$, while the acceleration during each collision is modeled as a series of impulsive changes of velocities:
\begin{equation}
\label{eqn:vdot}
    \dot{\mathbf{v}}_p(t')=\sum_{j=1}^{N_\mathrm{coll}^p}\Delta\mathbf{v}_{jp}\delta(t',t_{jp}),
\end{equation}
with, for each particle $p$, $\Delta\mathbf{v}_{jp}$ the vector-velocity change after each collision, $N_\mathrm{coll}^p$ the number of elastic collisions, and $\{t_{jp}\}$ the series of collision times randomly distributed. $N_\mathrm{coll}^p$ follows a Poisson distribution with parameter $\nu_\mathrm{el}T_p$. In addition, to include the transition radiations associated to the appearance and disappearance of the electrons, an effective acceleration is introduced at the initial and final times, $\dot{\mathbf{v}}(0)=\mathbf{v}_{0p}\delta(t',0)$ and $\dot{\mathbf{v}}(T_p)=-\mathbf{v}_{N_\mathrm{coll}^p}\delta(t',T_p)$.

The radiated electric field is obtained by summing each acceleration process. To get at a compact expression, the position vectors $\mathbf{r}_p(t)$ are modeled as a series of straight-line motions during two collision events, $\mathbf{r}_p(t)=\mathbf{r}_{jp}+\mathbf{v}_{jp}(t-t_{jp})$, with the velocity vectors $\mathbf{v}_{jp}$ randomly distributed. This neglects the influence of the electromagnetic wave of the radar on the particle motions, which is reasonable as long as the strength of this wave is weak. Besides, to avoid heavy notations, the refraction index of air is temporarily set to 1 and no distinction is made between $\mathbf{q}\times(\mathbf{q}\times\mathbf{x})$ and $\mathbf{x}$ as long as $\mathbf{x}$ is randomly distributed. Then, once the time integrations are carried out, the spectral energy density reads as
\begin{eqnarray}
\label{eqn:spectrum}
    \mathcal{E}_\nu(\Omega)&=&\frac{e^2}{16\pi^3\epsilon_0c^3}\bigg\langle\bigg|\sum_{p=1}^N\bigg(\frac{2eU_T\mathbf{n}_\Omega}{mR_T}\sum_{j=0}^{N_\mathrm{coll}^p}\frac{\sin{\bigg(\bigg(2\pi\Delta\nu-\Delta\mathbf{k}\cdot\mathbf{v}_{jp}\bigg)\Delta t_{jp}/2\bigg)}}{2\pi\Delta\nu-\Delta\mathbf{k}\cdot\mathbf{v}_{jp}}\exp{\big(-i\Delta\mathbf{k}\cdot\mathbf{R}_{0p}-i\varphi_{jp}\big)} \nonumber\\
    &+& \sum_{j=1}^{N_\mathrm{coll}^p} \Delta\mathbf{v}_{jp}\exp{(-i2\pi\nu t_{jp}-i\mathbf{k}\cdot(\mathbf{R}_{0p}-\mathbf{r}_{p}(t_{jp})))}-\mathbf{v}_{N_\mathrm{coll}^p,p}\exp{(-i2\pi\nu T_p)}+\mathbf{v}_{0p}\bigg)\bigg|^2\bigg\rangle,
\end{eqnarray}
with the phase factor $\varphi_{jp}=\big(2\pi\Delta\nu-\Delta\mathbf{k}\cdot\mathbf{v}_{jp}\big)\big(t_{jp}+\Delta t_{jp}/2\big)$ and the unit vector $\mathbf{n}_\Omega=\mathbf{q}\times(\mathbf{q}\times\mathbf{n}_T)$ specifying the direction of the electric field of the Thomson-scattered wave. In terms of the coordinate system of~\cref{fig:geom-thomson},
\begin{equation}
\label{eqn:nomega}
    \left|\mathbf{n}_\Omega\right|^2=1-\left|\mathbf{q}\cdot\mathbf{n}_T\right|^2=1-\sin^2{\theta_s}\cos^2{(\phi_0-\phi)},
\end{equation}
and for a randomly polarized incident wave, this reduces to $\left|\mathbf{n}_\Omega\right|^2=(1+\cos^2{\theta_s})/2$.

Eq.~(\ref{eqn:spectrum}) encompasses in an explicit way all dependencies of the spectral energy density searched for in terms of the incoming frequency of the wave $\nu_{\mathrm{T}}$, the time period $T_p$ during which each electron is free, and the elastic collision rate $\nu_{\mathrm{el}}$. The first term corresponds to the Thomson scattering in the presence of collisions. It is smoothly peaked for frequencies around $\nu_{\mathrm{T}}$, with a width proportional to $\langle T_p^{-2}\rangle$. For a volume large enough compared to $(2\pi/\nu_{\mathrm{T}})^3$, the presence of the term $\Delta\mathbf{k}\cdot\mathbf{R}_{0p}$ in the phase implies destructive interference between electrons located in $\mathbf{R}_{0}$ and in $\mathbf{R}_{0}^\prime= \mathbf{R}_{0}+\pi/\Delta\mathbf{k}$. For a plasma with uniform density, the energy effectively radiated from this scattering is thus entirely due to density fluctuations that produce an unbalanced number of electrons at different positions. The amplitude of such fluctuations being proportional to $\sqrt{N}$, the radiated energy is proportional to $N$. The other terms correspond to the spontaneous radiations of the plasma through Bremsstrahlung and transitions. Due to their stochastic nature, they lead as well to incoherent radiations except for Bremsstrahlung for frequencies below the collision rate where interference effects between successive collisions of the same particle suppress the emission. Interference effects between the different types of radiations are weak.

\section{Perspectives for the Detection of Extensive Air Showers?}

The plasma left in the atmosphere after the passage of an EAS is very different from a plasma with uniform density: the number of electrons changes rapidly with the distance to the shower axis. In the wavelength range of interest for radar applications, the variation of the electron number over distances where the phase of the incoming wave changes by $\pi$ is large enough to avoid destructive interference. According to eq.~(\ref{eqn:spectrum}), this might offer the opportunity of producing a signal proportional to the square of the number of emitting particles, and thus of significantly increasing the signal due to the re-radiation of particles undergoing Thomson scattering. 

Estimates of the spectral intensity require to model the 2-pt correlation function of the electron velocities. The difficulty is to infer an expression for $\left\langle\left(n'\mathbf{v}(t')\right)\cdot\left(n''\mathbf{v}(t'')\right)\right\rangle$ that renders all the coherence effects. In addition, the case of a uniform-density plasma is irrelevant and the emission terms need to be plugged into a model of EAS developing in the atmosphere to estimate in the end the ``radar cross section'', tuning the best way the incoming wavelength and the frequency range of detection for maximising this cross section and thus estimate whether such a technique might offer some perspectives or not.

\bibliographystyle{JHEP}
\bibliography{biblio}


\end{document}